\documentclass[twocolumn,showpacs,preprintnumbers,amsmath,amssymb,aps]{revtex4}
\usepackage{mathrsfs}
\usepackage[hypertex,linkcolor=red]{hyperref}
\def\d{\operatorname{d}}\def\<{\langle}\def\>{\rangle}
\def\Tr{\operatorname{Tr}}

\def\Base{\set{B}}
\def\grp#1{{\mathbf #1}}
\def\set#1{{\sf
#1}}
\def\dim{\operatorname{dim}} 
\def\Cmplx{\mathbb
C}
\def\Reals{\mathbb R}
\def\spc#1{\mathcal{#1}}
\def\rep#1{{\sf #1}}
\def\Bnd#1{\mathcal{B}(#1)} 
\def\Bndd#1,#2{\mathcal{B}(#1,#2)}
\def\SU#1{\mathbb{SU}(#1)}
\def\Proof{\medskip\par\noindent{\bf Proof. }}
\def\qed{$\,\blacksquare$\par}
\newtheorem{lemma}{Lemma}
\newtheorem{theo}{Theorem}
\newtheorem{prop}{Proposition}
\begin{document}
\title{Optimal estimation of group transformations using entanglement}
\author{G.~Chiribella} \email{chiribella@unipv.it}
\author{G.~M.~D'Ariano} \email{dariano@unipv.it} \altaffiliation[Also
at ]{Center for Photonic Communication and Computing, Department of
Electrical and Computer Engineering, Northwestern University,
Evanston, IL 60208} \author{M.~F.~Sacchi} \email{msacchi@unipv.it}
\affiliation{QUIT Quantum Information Theory Group of the INFM,
Unit\`a di Pavia}
\homepage{http://www.qubit.it} \affiliation{Dipartimento di Fisica
``A. Volta'', via Bassi 6, I-27100 Pavia, Italy} \date{\today}
\begin{abstract}
We derive the optimal input states and the optimal quantum
measurements for estimating the unitary action of a given symmetry
group, showing how the optimal performance is obtained with a suitable
use of entanglement. Optimality is defined in a Bayesian sense, as
minimization of the average value of a given cost function. We
introduce a class of cost functions that generalizes the Holevo class
for phase estimation, and show that for states of the optimal form all
functions in such a class lead to the same optimal measurement. A
first application of the main result is the complete proof of the
optimal efficiency in the transmission of a Cartesian reference
frame. As a second application, we derive the optimal estimation of a
completely unknown two-qubit maximally entangled state, provided that
$N$ copies of the state are available. In the limit of large $N$, the
fidelity of the optimal estimation is shown to be $1-3/(4N)$.
\end{abstract}
\maketitle
\section{Introduction}
Many of the most surprising advantages offered by the new technology
of Quantum Information \cite{Nielsen} arise from the concept of quantum
entanglement. Computational speed-up \cite{shor,grover}, quantum
teleportation \cite{teleport} and dense coding \cite{dense}, secure
protocols in cryptography \cite{Ekert}, precision enhancement in
quantum measurements  \cite{DLP,LorReview} are just a short summary of
some of the main lines of research inspired by entanglement.

After a so promising list, it is natural to expect remarkable
improvements coming from entanglement also in the context of Quantum
Estimation Theory \cite{Helstrom,Holevo}, in particular in the typical
problem of estimating an unknown physical transformation drawn from a
given set. With a heuristic argument inspired by dense coding, we
could expect that the accuracy in the discrimination of a set of
quantum channels can be increased by letting them act locally on a fixed side
of a maximally entangled state. Even more, one is tempted to guess
that a maximally entangled state is the optimal input for the
estimation of an unknown black box. Even though these are both
reasonable conjectures, in general they turn out to be false: for
example, a maximally entangled input state is always useless---and
often suboptimal---for the discrimination of two unitary
transformations \cite{AcinUnitaries,DLP}. The question then arises: is
it really possible to make some general statement about the role of
entanglement in the optimal estimation of an unknown transformation?

In this paper we will answer this question in the \emph{covariant
case}, which corresponds to the estimation of unitary transformations
randomly picked out from a given representation of some group. To face
the problem, we will choose the Bayesian approach, assuming a uniform
\emph{a priori} distribution for the unknown group parameters, and
defining optimality as the minimization of the average value of a
given cost function. Within the Bayesian framework, some results about
the optimality of maximally entangled states have been presented in
\cite{AJV,DLP}. Other results in the same direction have been derived
in \cite{Fujiwara,Ballester} within a different approach based on
quantum Cram\'er-Rao bound. However, all the mentioned results are
limited to particular cases, and their extension to arbitrary
representations of arbitrary groups is not straightforward.

Another nontrivial question is: which kind of entanglement is really
useful for the estimation of group transformations? In
Ref. \cite{AJV}, it has been considered the estimation of unitary
transformations $U_g$ in $\SU d$ in the form $U_g^{\otimes N}$,
corresponding to $N$ copies of the same unknown black box. The result
is that the optimal performance can be attained by entangling the $N$
$d$-level systems that undergo the unknown transformation with another
set of $N$ $d$-level systems playing the role of a \emph{reference
system}.  However, as pointed out in Ref. \cite{refframe}, the
entanglement with an additional set of $N$ reference systems actually
is not needed: what really matters is something more subtle, namely
the entanglement between spaces where the action of the group is
irreducible and spaces where the action of the group is trivial. In
the language of group theory, what is needed is maximal entanglement
between \emph{representation spaces} and \emph{multiplicity
spaces}. This kind of entanglement can be obtained not only by adding
an external reference system as in \cite{AJV}, but also via the use of
the multiple equivalent representations that appear in the
Clebsch-Gordan decomposition of the representation $\{U_g^{\otimes
N}\}$.

The concept of entanglement between representation spaces and
multiplicity spaces will be the protagonist of this paper. In the
following, we derive the optimal scheme for estimating an unknown
group transformation, showing how this kind of entanglement allows to
achieve the ultimate precision limits allowed by Quantum Mechanics.
To do this, we introduce a class of cost functions that generalize the
well known Holevo class for phase estimation \cite{Holevo}, and show
that all functions in such a class lead to the same optimal
measurement. We give also an explicit expression for the average cost
so that the optimization of the estimation scheme is reduced to a
simple eigenvalue problem.

In Sec. \ref{Tools}, before starting the analysis about optimal
estimation strategies, we introduce the notation (\ref{Notation}) and
some group theoretical tools (\ref{ElGrpTheory}) that will be
exploited throughout the paper. In Sec. \ref{SecOptEstimation}, we
present the problem of estimating an unknown group transformation
(\ref{TheProblem}), introducing a generalization to arbitrary groups
of the Holevo class of cost functions (\ref{GenHolevoClass}). The
optimal input states are then derived (\ref{SecOptInputStates}), and
the entanglement between representation and multiplicity spaces is
recognized to be the basic resource for an optimal estimation
strategy. In order to find the optimal measurement for the estimation
of a group transformation, we show in Sec. \ref{CovProperties} how the
special form of the optimal input states reflects on the covariance
properties of the optimal measurement. Exploiting this analysis, we
will show in \ref{SecOptPovm} that, for input states of the optimal
form, all cost functions in the generalized Holevo class lead to the
same optimal measurement. Finally, Sec. \ref{Applications} is devoted
to applications of the general results. A first application
(\ref{RefframeProof}) is the optimality proof of the protocol
\cite{refframe} for the absolute alignment of two reference
frames. As a second application, we derive (\ref{EntEstimation}) the
optimal estimation of a completely unknown two-qubit maximally
entangled state with $N$ identical copies of the state. Section V
concludes the paper, while the most technical proofs are provided in
the Appendix.
\section{Theoretical tools}\label{Tools}

\subsection{Notation for bipartite states}\label{Notation}
A simple notation can be introduced to deal with bipartite
states. Given two Hilbert spaces $\spc H_A$ and $\spc H_B$, and fixed
two orthonormal bases $\mathcal{B}_A=\{|\phi_n\>~|~ n=1, \dots, d_A\}$
and $\mathcal{B}_B=\{|\psi_n\>~|~ n=1, \dots, d_B\}$ for $\spc H_A$
and $\spc H_B$ respectively, it is possible to associate in a one to
one way any vector $|C\>\!\> \in \spc H_A \otimes \spc H_B$ with an
operator $C \in \Bndd {\spc H_B}, {\spc H_A} $ via the relation
\cite{PLA}
\begin{equation}\label{DoubleKet}
|C\>\!\>=\sum_{m,n}~ \<\phi_m|C|\psi_n\>~ |\phi_m\>|\psi_n\>~.
\end{equation}
With this notation, one has the simple relations
\begin{equation}\label{DoubleKetProduct}
\<\!\<C|D\>\!\>=\Tr[C^{\dag}D]\qquad 
\end{equation}
and
\begin{equation}\label{DoubleKetProperty}
A \otimes B~|C\>\!\>= |ACB^T\>\!\>~,
\end{equation}
for any $A \in \Bnd {\spc H_A}$ and $ B\in \Bnd{\spc H_B}$, where
transposition $T$ is defined with respect to the fixed bases.  Such
relations allow to greatly simplify the calculation involving
entangled states, and will be extensively used throughout the paper.
\subsection{Elements of group theory}\label{ElGrpTheory}
Here we recall some simple tools of group theory \cite{groups} that
will be exploited throughout the paper.

Suppose we are given a Hilbert space $\spc H$ and a unitary
representation $\rep R (\grp G)= \{U_g\in \Bnd {\spc H}~|~g \in \grp
G\}$ of a compact Lie group $\grp G$.  The Hilbert space can be
decomposed into orthogonal subspaces in the following way
\begin{equation}\label{SpaceDecomp}
\spc H \equiv \bigoplus_{\mu \in \set S}~ \spc H_{\mu} \otimes
\Cmplx^{m_{\mu}}~,
\end{equation}
where the sum runs over the set of irreducible representations of
$\grp G$ that appear in the Clebsch-Gordan decomposition of $\rep
R(\grp G)$.  The action of the group is irreducible in each
\emph{representation space} $\spc H_{\mu}$, while it is trivial in the
\emph{multiplicity space} $\Cmplx^{m_{\mu}}$, namely
\begin{equation}\label{RepDecomp}
U_g \equiv \bigoplus_{\mu \in \set S}~U_g^{\mu} \otimes \openone_{m_{\mu}}~,
\end{equation}
 $\openone_d$ denoting the identity in a $d-$dimensional Hilbert
space.  The projection $\Pi_{\mu}$ onto the subspace $\spc H_{\mu}
\otimes \Cmplx^{m_{\mu}}$ is given by the integral formula
\begin{equation}\label{IntegralProject}
\Pi_{\mu}=d_{\mu}~\int \d g ~ \chi^{\mu*}(g)~U_g~,
\end{equation}
where $\d g$ denotes the normalized invariant Haar measure ($\d g=\d
(kg)= \d (gk)$ for any $k,g \in \grp G$), $d_{\mu}\equiv \dim (\spc
H_{\mu})$, and $\chi^{\mu}(g)\equiv\Tr[U^{\mu}_g]$ is the character of
the irreducible representation $\mu$.  Note that here we are
considering $\grp G$ as a continuous group only for fixing notation,
nevertheless---here and all throughout the paper---$\grp G$ can have a
finite number of elements, say $|\grp G|$, and in this case we have
simply to replace integrals with sums and $\d g $ with $1/|\grp G|$.

Moreover, any operator $O \in \Bnd {\spc H}$ in the commutant of $\rep
R(\grp G)$---i.e. such that $[O,U_g]=0 \quad \forall g \in \grp
G$---has the form
\begin{equation}\label{CommutingOp}
O= \bigoplus_{\mu \in \set S}~ \openone_{d_{\mu}} \otimes O_{\mu}~,
\end{equation}
where $O_{\mu}$ is a $m_{\mu} \times m_{\mu}$ complex matrix.  In
 particular, the group average $\<A\>_{\grp G}\equiv \int \d g~ U_g A
 U_g^{\dag}$ of a given operator $A$ with respect to the invariant
 Haar measure is in the commutant of $\rep R(\grp G)$, and has the
 form:
\begin{equation}\label{AveOp}
\<A\>_{\grp G}=\bigoplus_{\mu \in \set S}~ \openone_{d_{\mu}} \otimes
\frac{1}{d_{\mu}}~ \Tr_{\spc H_{\mu}}[A]~,
\end{equation}  
where $\Tr_{\spc H_{\mu}}[A]$ is a short notation for $\Tr_{\spc
H_{\mu}}[\Pi_\mu~ A~\Pi_\mu]$, $\Pi_\mu$ being the projection onto
$\spc H_{\mu} \otimes \Cmplx^{m_{\mu}}$.  Here and throughout the
paper we assume the normalization of the Haar measure: $\int_{\grp G}
\d g=1$.

\bigskip
{\bf Remark I:} \emph{entanglement between representation spaces and
multiplicity spaces.}\\ The choice of an orthonormal basis
$\Base^{\mu}~=~\{|\phi_n^{\mu}\>~\in~\Cmplx^{m_{\mu}}~|~n=1, \dots
,m_{\mu}\}$ for a multiplicity space fixes a particular decomposition
of the Hilbert space as a direct sum of irreducible subspaces:
\begin{equation}\label{DirectSum}
\spc H_{\mu} \otimes \Cmplx^{m_{\mu}} = \oplus_{n=1}^{m_{\mu}}~ \spc
H^{\mu}_n~,
\end{equation}
where $\spc H_n^{\mu} \equiv \spc H_{\mu} \otimes |\phi_n^{\mu}\>$.
In this picture, it is clear that $m_\mu$ is the number of different
irreducible subspaces carrying the same representation $\mu$, each of
them having dimension $d_{\mu}$.  Moreover, with respect to the
decomposition (\ref{SpaceDecomp}), any pure state $|\Psi\> \in \spc H$
can be written as
\begin{equation}\label{StateDecomp}
|\Psi\>= \bigoplus_{\mu \in \set S}~ c_{\mu}~|\Psi_{\mu}\>\!\>~,
\end{equation} 
where $|\Psi_{\mu}\>\!\>$ is a bipartite state in $\spc H_{\mu}
\otimes \Cmplx^{m_{\mu}}$ and $\sum_{\mu \in \set
S}~|c_{\mu}|^2=1$. With respect to the direct sum decomposition
(\ref{DirectSum}), the Schmidt number of such a state is the minimum
number of subspaces carrying the same representation $\mu \in \set S$
that are needed to decompose $|\Psi\>$.
 
\bigskip
{\bf Remark II:} \emph{maximum number of equivalent representations in
the decomposition of a pure state.}\\ The Schmidt number of any
bipartite state $|\Psi_{\mu}\>\!\>~\in~\spc
H_{\mu}~\otimes~\Cmplx^{m_{\mu}}$ is always less then or equal to
$k_{\mu}= \min \{d_{\mu}, m_{\mu}\}$.  This means that any pure state
can be decomposed using \emph{no more} than $k_{\mu}$ irreducible
subspaces carrying the same representation $\mu \in \set S$.

\section{Optimal estimation of group transformations}\label{SecOptEstimation} 
\subsection{Background problem}\label{TheProblem}
Suppose we are given a black box that performs on a system
$\mathcal{S}$ an unknown unitary transformation $U_g$ randomly drawn
from a group representation $\rep R ({\grp G})$.  In order to estimate
the transformation $U_g$, we can prepare the system in an input state
$\rho^{\mathcal{S}}$, send it through the black box, and try to
estimate the parameter $g$ from the output state
\begin{equation}
\rho^{\mathcal{S}}_g\equiv U_g \rho^{\mathcal{S}} U_g^{\dag}~.
\end{equation} More generally, we can also exploit an
additional reference system $\mathcal{R}$ and prepare an entangled
state $\rho^{\mathcal{SR}}$, so that the output state becomes
\begin{equation}
\rho^{\mathcal{SR}}_g\equiv (U_g \otimes \openone_{\mathcal{R}})
~\rho^{\mathcal{SR}}~ (U_g^{\dag} \otimes \openone_{\mathcal{R}})~.
\end{equation}
Our task is to find the best input states and the best estimation
strategies allowed by Quantum Mechanics to determine the parameter
$g$. Since we are interested in ultimate in-principle limits, we
assume complete freedom in preparing any physical state and in
realizing any quantum measurement.  This means that we are allowed to
choose the state $\rho^{\mathcal{SR}}$ with minimal stability group,
reducing the set of unitaries that are not discriminable to those that
differ just by a phase factor. Therefore, the stability group can be
only a (nontrivial) center for $\grp G$, made of multiples of the
identity, corresponding to (a subgroup of) $U(1)$. The quotient group
is then a group itself, and in the following we will use the same
symbol $\grp G$ for such a quotient group. Notice that the requirement
of central stability group $U(1)$ is satisfied by choosing the state
$\rho^{\mathcal{SR}}$ as pure, and with maximal Schmidt number.

The most general estimating strategy allowed by quantum mechanics,
including both quantum measurements and classical data processing, can
be described by a Positive Operator Valued Measure (POVM) $M$ that
associates to any estimate ${\hat g} \in \grp G$ a positive
semidefinite operator $M({\hat g})$, satisfying the normalization
condition
\begin{equation}\label{Norm}
\int_{\grp G}~ \d g~ M(g)= \openone~.
\end{equation} 
The probability density of the estimate ${\hat g}$ in the state
$\rho_g$ is given by the usual Born rule:
\begin{equation}\label{ProbDens}
p({\hat g}|g)= \Tr[\rho_g ~M({\hat g})]~.
\end{equation}

In this paper, the estimation problem will be faced in the Bayesian
setting with prior uniform probability density $\d g$, and the optimal
estimation will be defined as the one that minimizes the average value
of a given cost function $c({\hat g},g)$ that associates to any
estimate ${\hat g}$ a cost which increases versus the ``distance'' of
$\hat g$ from the true value $g$.  The average of the cost function
over the prior and the conditional probability distributions will be
given by
\begin{equation}\label{AveC}
\<c\>= \int \d g~\int \d {\hat g}~ c({\hat g},g)~ p({\hat g}|g)~.
\end{equation}

\subsection{A generalized Holevo class of cost functions}\label{GenHolevoClass}
We will make two assumptions on the form of the cost function $c({\hat
g},g)$.

\emph{First assumption.} We require $c$ to be group invariant, namely
\begin{equation}\label{LeftInv}
c({\hat g},g)=c(k{\hat g},kg) \qquad \forall {\hat g}, g, k \in \grp G
\end{equation}
(left-invariance), and
\begin{equation}\label{RightInv}
c({\hat g},g)=c(gk,{\hat g}k) \qquad \forall {\hat g}, g, k \in \grp G
\end{equation}
(right-invariance).
By using Fourier analysis, one can prove (see Appendix) that this assumption is equivalent to the
expansion 
\begin{equation}\label{OurC}
c({\hat g},g)= \sum_{\sigma}~ a_{\sigma}~ \chi^{\sigma*}({\hat g}g^{-1})~,
\end{equation}
where $\chi^{\sigma}(g)\equiv \Tr[U^{\sigma}(g)]$ is the character of the
irreducible representation $\sigma$, and the coefficients $a_{\sigma}$ 
satisfy the identity $a_{\sigma}^*=a_{\sigma *} \quad \forall \sigma$, in order to have a real cost function.

\emph{Second assumption.} We require all nonzero coefficients
$a_{\sigma}$ in the expression (\ref{OurC}) to be negative, with the
only exception of the coefficient $a_{\sigma_0}$ corresponding to the
trivial representation $U^{\sigma_0}(g)=1
\quad \forall g$, which is allowed to be positive (the $\sigma_0$ term just adds a trivial constant
to the cost function, since $\chi^{\sigma_0}(g)=1 \quad \forall g$).

The class of functions that satisfy our two assumptions is a direct
generalization of the class of cost functions introduced by Holevo for
the estimation of phase shifts \cite{Holevo}. In fact, such functions
have the form
\begin{equation}
c(\hat \phi -\phi)=\sum_{k \in \mathbb{Z}}~a_k~ e^{-ik (\hat \phi -\phi)}~,
\end{equation}
where $a_k \le 0$ for any $k \neq 0$, and $e^{ik\phi}$ is the character of the unidimensional representation labeled by $k$.

\subsection{Optimal choice of the input state}\label{SecOptInputStates}
Since the average cost (\ref{AveC}) is a linear functional of the
input state $\rho$, in the optimization problem we can restrict
attention to \emph{pure} input states $\rho=|\Psi\>\<\Psi|$. Then the
problem becomes equivalent to the optimal discrimination problem of
states in the orbit
\begin{equation}\label{Orbit}
\mathcal{O}=\left\{|\Psi_g\>\equiv U_g~ |\Psi\>~|~ g \in \grp G\right\}
\end{equation}
generated from $|\Psi\>$ by the action of the representation $\rep R(\grp G)$.

Let's consider the Clebsch-Gordan decomposition (\ref{RepDecomp}) of
the unitaries $U_g$. From now on we will assume the algebraic
condition
\begin{equation}\label{EnoughMult}
m_{\mu}= d_{\mu} \qquad \forall \mu \in \set S~.
\end{equation} 
\begin{lemma} The assumption (\ref{EnoughMult}) can 
be done without any loss of generality.
\end{lemma} 
\Proof Suppose $d_{\mu} >m_{\mu}$ for some representation $\mu$. In
this case, we can introduce a reference system $\mathcal{R}$ whose
dimension is
\begin{equation}
d_{\mathcal{R}}\geq \max_{\mu \in \set S}
\left\{\frac{d_{\mu}}{m_{\mu}}\right\}~,
\end{equation}
and replace $U_g$ with its extension $U_g'=U_g \otimes
\openone_{\mathcal{R}}$, acting in the tensor product Hilbert space
$\spc H \otimes \spc H_{\mathcal{R}}$. In this way, $U_g'$ will
satisfy the condition \mbox{$m'_{\mu}\equiv m_{\mu}\times
d_{\mathcal{R}}\geq d_{\mu} \quad \forall \mu$}.  On the other hand,
as already mentioned at the end of Sec. \ref{ElGrpTheory}, any pure
state $|\Psi\>$ can be decomposed in the form (\ref{StateDecomp}) with
no more than \mbox{$k_{\mu}=\min \{d_{\mu},m_{\mu}\}$} irreducible
subspaces for any $\mu$.  Therefore, we can switch our attention from
the whole Hilbert space $\spc H\otimes \spc
H_{\mathcal{R}}=\bigoplus_{\mu}~\spc H_{\mu} \otimes
\Cmplx^{m'_{\mu}}$ to the invariant subspace \mbox{$\spc H'\equiv
\bigoplus_{\mu}~\spc H_{\mu} \otimes \Cmplx^{d_{\mu}}$}, which
contains the input state $|\Psi\>$ along with its orbit (\ref{Orbit}).
In other words, without loss of generality we can always consider an
input state in the Hilbert space
\begin{equation}\label{PsiSpace}
\spc H'=\bigoplus_{\mu}~\spc H_{\mu} \otimes \Cmplx^{d_{\mu}}~,
\end{equation}
which can be thought as embedded in a larger Hilbert space $\spc H
 \otimes \spc H_{\mathcal{R}}$.  \qed {\bf Remark.} The need of adding
 an external reference system $\mathcal{R}$ arises only in the case
 when $d_{\mu}>m_{\mu}$ for some irreducible representation $\mu$.  In
 fact, the role of the reference system is simply to increase the
 number of equivalent representations until the extended Hilbert space
 $\spc H \otimes \spc H_{\mathcal{R}}$ reaches the threshold
 $m_{\mu}\geq d_{\mu} \quad \forall \mu$. This observation allows to
 greatly reduce the dimension of the reference system with respect to
 the customary estimation schemes inspired by dense coding, with a
 reference system $\spc H_{\mathcal{R}}$ having the same dimension of
 $\spc H$.

Now we show that the best input state $|\Psi\>$ for estimating the
group transformation of an unknown black box is a state of the form
(\ref{StateDecomp}), with each $|\Psi_{\mu}\>\!\>$ maximally
entangled, namely
\begin{equation}
|\Psi_{\mu}\>\!\>= \frac{1}{\sqrt{d_{\mu}}}~ \sum_{n=1}^{d_{\mu}}~
|\psi_n^{\mu}\>|\phi_n^{\mu}\>~,
\end{equation}
 $\Base_A^{\mu}=\{|\psi_n^{\mu}\>~|~ n=1, \dots , d_{\mu}\}$ and
 $\Base_B^{\mu}=\{|\phi_n^{\mu}\>~|~ n=1, \dots , d_{\mu}\}$ being
 Schmidt bases for $\spc H_{\mu}$ and $\Cmplx^{d_{\mu}}$
 respectively. Exploiting the notation (\ref{DoubleKet})---with fixed
 bases $\Base_A^{\mu}$ and $\Base_B^{\mu}$---the optimal input state
 $|\Psi\>$ must have the form
\begin{equation}\label{PsiEntangled}
|\Psi\>= \bigoplus_{\mu \in \set S}~ \frac{c_{\mu}}{\sqrt{d_{\mu}}}~ |W_{\mu}\>\!\>~,
\end{equation} 
with $W_{\mu}\equiv \sum_n~ |\psi^{\mu}_n\>\<\phi^{\mu}_n|$ unitary
operators.
\begin{theo}[optimal input states]\label{Theo1} With a suitable choice of the coefficients $\{c_{\mu}\}$, any input state of the form (\ref{PsiEntangled}) achieves the minimum average cost.
\end{theo}
Suppose that the minimum cost $\<c\>^{Opt}$ is achieved by the input
state $|\Phi\>=\bigoplus_{\mu}~c_{\mu}~|\Phi_{\mu}\>\!\>$ along with
the estimation strategy described by the POVM $M(g)$. The operator
$K_h\equiv \bigoplus_{\mu}~ \openone_{\mu} \otimes
\sqrt{d_{\mu}}\left( W_{\mu}^{\dag} U_h^{\mu}\Phi_{\mu}\right)^T$
converts the orbit of an input state (\ref{PsiEntangled}) into the
orbit of the optimal input state $|\Phi\>$, since using identity
(\ref{DoubleKetProperty}), we have
\begin{equation}
K_h~|\Psi_g\>=|\Phi_{gh}\>~,
\end{equation}
where $|\Psi_g\>=U_g|\Psi\>$ and $|\Phi_g\>=U_g|\Phi\>$.  Consider now
the POVM $M'(g)\equiv \int \d h ~K_h^{\dag}~M(gh)~K_h$. The POVM
$M'(g)$ is normalized, since
\begin{eqnarray*}
\int \d g~M'(g)&=&\int \d g \int \d h\, K_h^{\dag}~M(gh)~K_h\\
&=& \int \d h~ K_h^{\dag}K_h\\
&=&\openone~,
\end{eqnarray*}
where we exchanged integrals over $g$ and $h$, used invariance of the
Haar measure $\d g$, and finally used Eq. (\ref{AveOp}) and the
normalization of bipartite states $|\Phi_{\mu}\>\!\>$ in the form
$\Tr[\Phi_{\mu}^{\dag}\Phi_{\mu}]=1$.  A state $|\Psi\>$ of the form
(\ref{PsiEntangled}) along with the POVM $M'(g)$ achieves the minimum
cost. In fact,
\begin{eqnarray*}
\<c\>&=& \int \d g \int \d \hat g~ c(\hat g,g)~ \<\Psi_g|~M'(\hat g)~|\Psi_g\>\\&=& \int \d g \int \d \hat g \int \d h~ c(\hat g ,g)~\<\Phi_{gh}|~M(\hat g h)~|\Phi_{gh}\>~~~~~\\
&=& \int \d g \int \d \hat g \int \d h~  c(\hat g h,gh) \<\Phi_{gh}|~M(\hat
g h)~|\Phi_{gh}\>~~~~~~\\ &=& \int \d k \int \d \hat k~ c(\hat k, k)~\<\Phi_k|~M(\hat k)~|\Phi_k\>\\&=&\<c\>^{Opt}~,
\end{eqnarray*}  
where we used right-invariance of both cost function and Haar measure. \qed
\subsection{Covariance properties of the estimating POVM}\label{CovProperties}
Since the whole orbit (\ref{Orbit}) is generated from the input state
$|\Psi\>$ by the action $\rep R(\grp G)$ of the group, there is no
loss of generality in restricting attention to estimating POVM of the
covariant form \cite{Holevo}
\begin{equation}\label{CovPovm}
M(g)=U_g~\Xi~U_g^{\dag}
\end{equation} 
with $\Xi$ a suitable positive operator satisfying the normalization
condition (\ref{Norm}). A covariant POVM yields a left-invariant
probability distribution, namely $p(k\hat{g}|kg)=p(\hat{g}|g) \quad
\forall k, \hat{g}, g \in \grp G$. Using both the left-invariance of
the probability distribution and of the cost function, the average
cost (\ref{AveC}) can be written as
\begin{equation}\label{AveCBis}
\<c\>=\int \d g~ c(g,e)~ p(g|e)~
\end{equation}
where $e$ is the identity element of the group $\grp G$.  

For superpositions of maximally entangled states as in
Eq. (\ref{PsiEntangled}), the orbit $\mathcal{O}$ enjoys an additional
symmetry that reflects on an additional covariance property of the
POVM.  In fact, using the decomposition (\ref{RepDecomp}) and the
identity (\ref{DoubleKetProperty}), we can note that
\begin{eqnarray*}
|\Psi_g\>&=& U_g~|\Psi\>\\ &=&\bigoplus_{\mu \in \set
  S}~\frac{c_{\mu}}{\sqrt{d_{\mu}}}~(U_g^{\mu} \otimes
  \openone_{\mu})~|W_{\mu}\>\!\>\\ &=& \bigoplus_{\mu \in \set S}~
  \frac{c_{\mu}}{\sqrt{d_{\mu}}}~[\openone_{\mu} \otimes
  (W_{\mu}^{\dag}U_g^{\mu}W_{\mu})^T]~|W_{\mu}\>\!\>\\ &=&
  V_g^{\dag}~|\Psi\> \qquad \forall g \in \grp G~,
\end{eqnarray*}
where 
\begin{equation}\label{V}
V_g\equiv \oplus_{\mu \in \set S}~ (\openone_{\mu} \otimes
(W_{\mu}^{\dag}U_g^{\mu}W_{\mu})^* )~,
\end{equation}
is an element of a new unitary representation $\rep R'(\grp G)$ of the
group $\grp G$. Notice that the two representations $\rep R(\grp G)$
and $\rep R'(\grp G)$ commute among themselves.  Then, the following
Lemma holds:
\begin{lemma}\label{BiCovPOVM}
There is no loss of generality in assuming a covariant POVM $M(g)= U_g
~\Xi~ U_g^{\dag}$ with
\begin{equation}\label{XiComm}
[\Xi, U_gV_g]=0 \qquad \forall g \in \grp G~,
\end{equation}
where $U_g$ and $V_g$ are given in Eqs. (\ref{RepDecomp}) and
(\ref{V}), respectively.
\end{lemma} \Proof For any possible POVM $N(g)$
there is a covariant POVM with the above property and with the same
average cost.  In fact, the group average
\begin{equation}
M(g)=\int \d k \int \d h~
U_k^{\dag} V_h^{\dag}~ N(kgh^{-1})~ V_h U_k 
\end{equation} is covariant---namely
$M(g)=U_g~\Xi~ U_g^{\dag}$ with $\Xi=M(e)$---and satisfies the
required commutation relation (\ref{XiComm}). Both properties follow
simply from the invariance of the Haar measure. To prove that the cost
of the covariant POVM $M(g)$ is the same as the cost of $N(g)$ we use
the property
\begin{equation}
U_k V_h ~|\Psi_g\>=|\Psi_{kgh^{-1}}\> \qquad \forall k, h, g \in \grp
G
\end{equation} 
of the states generated from the input (\ref{PsiEntangled}). In this way,
\begin{eqnarray*}
\<c\>_M&\equiv&\int \d g\int \d \hat g ~c(\hat g,g)~\<\Psi_g|~M(\hat
g)~|\Psi_g\>\\ &=& \int \d g \int \d \hat g \int \d k \int \d h~
c(\hat g ,g)
~\times \\ && \qquad \qquad \qquad \times
\<\Psi_{kgh^{-1}}|~N(k \hat g h^{-1})~|\Psi_{kgh^{-1}}\>\\ &=& \int \d
g \int \d \hat g\int \d k \int \d h~ c(k \hat g
h^{-1},kgh^{-1})
~\times \\ && \qquad \qquad \qquad \times
\<\Psi_{kgh^{-1}}|~N(k \hat g h^{-1})~|\Psi_{kgh^{-1}}\>\\ &=&\int \d
r \int \d \hat r ~c(\hat r,r)~\<\Psi_r|~N(\hat r)~|\Psi_r\>\\ &\equiv&
\<c\>_N~,
\end{eqnarray*}
where we used the left- and right-invariance of the cost function
$c(\hat g,g)$.  \qed

Let's diagonalize the operator $\Xi$ and express its (non-normalized)
eigenvectors in the decomposition (\ref{SpaceDecomp}):
\begin{eqnarray}
\nonumber \Xi &=& \sum_{i=1}^r~ |\eta^i\>\<\eta^i|\\
\label{XiDecomp}
&=& \sum_i~ \bigoplus_{\mu,\nu}~ \sqrt{d_{\mu}
d_{\nu}}~|\eta^i_{\mu}\>\!\>\<\!\<\eta^i_{\nu}|~,
\end{eqnarray}
where the factor $\sqrt{d_{\mu}}$ has been inserted just for later
convenience.
\begin{lemma}
Any covariant POVM $M(g)=U_g ~\Xi~ U_g^{\dag}$ with the commutation
property (\ref{XiComm}) must satisfy the two relations:
\begin{equation}\label{Norm1}
\sum_i~\eta^{i\dag}_{\mu}\eta^{i}_{\mu} = \openone_{\mu} \qquad
\forall \mu \in \set S~,
\end{equation}
and
\begin{equation}\label{Norm2}
\sum_i~ \eta^i_{\mu} \eta^{i\dag}_{\mu}= \openone_{\mu} \qquad \forall \mu \in \set S~.
\end{equation}  
\end{lemma}
\Proof The normalization (\ref{Norm}) becomes
\begin{equation}\label{XiNorm}
\<\Xi\>_{\grp G}\doteq\int \d g ~ U_g ~\Xi~ U_g^{\dag}= \openone~.
\end{equation}  The group average $\<\Xi\>_{\grp G}$ can be expressed using Eq. (\ref{AveOp}). In this way,  Eq. (\ref{XiNorm}) becomes
\begin{equation}\label{TrNorm}
\frac{1}{d_{\mu}} \Tr_{\spc H_{\mu}}[\Xi]= \openone_{\mu} \qquad
\forall \mu \in \set S~.
\end{equation}
By explicit computation,
\begin{eqnarray*}
\frac{1}{d_{\mu}} \Tr_{\spc H_{\mu}} \left[ \Xi \right]&=& \sum_i~
\Tr_{\spc H_{\mu}} \left[
|\eta^i_{\mu}\>\!\>\<\!\<\eta^i_{\mu}|~\right]\\ &=& \sum_i~
\eta^{iT}_{\mu} ~\Tr_{\spc H_{\mu}} \left[~
|\openone_{\mu}\>\!\>\<\!\<\openone_{\mu}|~\right]~ \eta^{i*}_{\mu}\\
&=& \sum_i~ \eta^{iT}_{\mu}\eta^{i*}_{\mu}~.
\end{eqnarray*}
Substituting this expression in (\ref{TrNorm}) and taking the complex
conjugate we get (\ref{Norm1}).  Moreover, using the commutation
relation (\ref{XiComm}), we can transform the group average with
respect to $\rep R(\grp G)$ in a group average with respect to $\rep
R'(\grp G)$, namely
\begin{eqnarray*}\<\Xi\>_{\grp G}&=&\int \d g~ U_g~ (U_g^{\dag}V_g^{\dag} ~\Xi~ U_g V_g)~U_g^{\dag}\\
&=&\int \d g~ V_g^{\dag} ~\Xi~ V_g~.
\end{eqnarray*}
In this way, using Eq. (\ref{V}), Eq. (\ref{Norm2}) can be proved
along the same lines used to prove Eq. (\ref{Norm1}). \qed

\subsection{The optimal POVM}\label{SecOptPovm}
We are now able to find the optimal covariant POVM for the estimation
of group transformation with superpositions of maximally entangled
states.
\begin{theo}[optimal POVM]\label{Theo2}
In the estimation of the states in the orbit $\mathcal{O}$ generated
from the input state
\begin{equation}
|\Psi\>=\bigoplus_{\mu \in \set S}~
 \frac{c_{\mu}}{\sqrt{d_{\mu}}}~|W_{\mu}\>\!\>~,
\end{equation}
where $W_{\mu}$ are unitary operators, the covariant POVM given by
$\Xi= |\eta\>\<\eta|$ with
\begin{equation}\label{OptEta}
|\eta\>=\bigoplus_{\mu \in \set S}~\sqrt{d_{\mu}}~e^{i\arg (c_{\mu})}~ |W_{\mu}\>\!\>
\end{equation}
is optimal for any cost function $c({\hat g},g)$ of the form
\begin{equation}\label{OurCBis}
c({\hat g},g)=\sum_{\sigma}~a_{\sigma}~\chi^{\sigma *}({\hat g}g^{-1})~,
\end{equation}
with $a_{\sigma}\leq 0 \qquad \forall \sigma \not = \sigma_0$.\\ The
average cost corresponding to the optimal estimation strategy is
\begin{equation}\label{OptCost}
\<c\>^{Opt}= a_{\sigma_0}+\sum_{\mu ,\nu}~ |c_{\mu}| ~C_{\mu\nu}~
|c_{\nu}|~,
\end{equation}
where 
\begin{equation}\label{Matrix}
C_{\mu\nu}\equiv \sum_{\sigma\not= \sigma_0}~
a_{\sigma}~m_{\sigma}^{(\mu\nu)}~,
\end{equation}
$m^{(\mu\nu)}_{\sigma}$ being the multiplicity of the irreducible
representation $\sigma$ in the Clebsch-Gordan series of the tensor
product $U^{\mu}_g \otimes U^{\nu*}_g$.
 
\end{theo} 
\Proof We will show that Eq. (\ref{OptCost}) gives a lower bound for
the average cost, and that the POVM $\Xi= |\eta\>\<\eta|$ with
$|\eta\>$ given by Eq. (\ref{OptEta}) achieves this bound.  By using
identities (\ref{DoubleKetProduct}) and (\ref{DoubleKetProperty}), and
the form (\ref{XiDecomp}) for the operator $\Xi$, Eq. (\ref{AveCBis})
becomes
\begin{eqnarray*} \<c\>&=&\int \d g ~c(g,e) 
\times\\ && \nonumber \times 
\sum_i~
\sum_{\mu,\nu}~c_{\mu}^*c_{\nu}~\Tr\left[W_{\mu}^{\dag}~U_{g}^{\mu}~
\eta^i_{\mu} \otimes W_{\nu}^T ~U_g^{\nu *}~ \eta^{i*}_{\nu}\right]~.
\end{eqnarray*}
Let's expand $c(g,e)$ as in (\ref{OurCBis}). Subtracting from the
average cost $\<c\>$ the constant term $a_{\sigma_0}$, which is not
relevant for the optimization, we get
\begin{eqnarray*}
\<c\>-a_{\sigma_0}&=& \sum_i \sum_{\mu
,\nu}~c_{\mu}^*c_{\nu}
~\times \\ \nonumber &&\times~
\sum_{\sigma\not=\sigma_0}~ \frac{a_{\sigma}}{d_{\sigma}}
~\Tr\left[\Pi^{(\mu\nu)}_{\sigma}~(\eta^i_{\mu}W_{\mu}^{\dag}\otimes
\eta^{i*}_{\nu}~W_{\nu}^T)\right]~,
\end{eqnarray*}
where we defined
\begin{equation}
\Pi^{(\mu\nu)}_{\sigma}\equiv d_{\sigma}~ \int \d g~
\chi^{\sigma*}(g)~ U_g^{\mu} \otimes U_g^{\nu*}~.
\end{equation}
According to (\ref{IntegralProject}), $\Pi^{(\mu\nu)}_{\sigma}$ is the
projection onto the direct sum of all the subspaces of $\spc H_{\mu}
\otimes \spc H_{\nu}$ that carry the irreducible representation
$\sigma$ in the tensor product $U^{\mu}_g \otimes U^{\nu*}_g$.
Clearly $\Pi_{\sigma}^{(\mu\nu)}$ is nonzero if and only if the
Clebsch-Gordan series of $U^{\mu}_g \otimes U^{\nu^*}_g$ contains
$\sigma$ with nonzero multiplicity $m_{\sigma}^{(\mu\nu)}$. Notice
also that
$\Tr[\Pi^{(\mu\nu)}_{\sigma}]=d_{\sigma}m_{\sigma}^{(\mu\nu)}$, by
definition of $\Pi^{(\mu\nu)}_{\sigma}$.

Denoting by $\sum_{\mu,\nu,\sigma}'$ the sum over $\mu,\nu$ and all
$\sigma$ except $\sigma_0$, the average cost can be bounded as follows
\begin{equation*}
\begin{split}
\<c\>-a_{\sigma_0}&\geq {\sum _{\mu,\nu,\sigma}}'
\ \frac{a_{\sigma}}{d_{\sigma}} \left| c_{\mu} c_{\nu}\sum_i \Tr\left[
\Pi_{\sigma}^{(\mu\nu)}(\eta_{\mu}^i~W_{\mu}^{\dag} \otimes
\eta_{\nu}^{i*}~W_{\nu}^T)\right] \right|~~~~~~~~\\ &\geq {\sum
_{\mu,\nu,\sigma}}' \ \frac{a_{\sigma}}{d_{\sigma}}
|c_{\mu}c_{\nu}|\sqrt{\left( \sum_i \Tr\left[\Pi_{\sigma}^{(\mu\nu)}
(\eta_{\mu}^{i}\eta_{\mu}^{i\dag} \otimes
\openone_{\nu})\right]\right)}~~~\\ \nonumber & \times\sqrt{\left(
\sum_j \Tr \left[\Pi_{\sigma}^{(\mu\nu)}(\openone_{\mu} \otimes
W_{\mu}^*\eta^{jT}_{\nu}\eta^{j*}_{\nu}W_{\mu}^T)\right] \right)}~,~~~
\end{split}
\end{equation*} 
since all $a_\sigma$ are nonpositive.  The second inequality follows
from Cauchy-Schwartz inequality with respect to the scalar product
$\<\mathbf{A},\mathbf{B}\> \equiv \sum_i ~\Tr
\left[A_i^{\dag}B_i\right]$, where we take
$A_i^{\dag}=\Pi^{(\mu\nu)}_{\sigma}(\eta^i_{\mu}W_\mu^\dag\otimes
\openone_{\nu})$ and $B_i=(\openone_{\mu}\otimes
\eta^{i*}_{\nu}W_\mu^T)~\Pi_{\sigma}^{(\mu\nu)}$.  Exploiting the
relations (\ref{Norm1}) and (\ref{Norm2}), and using that $\Tr\left[
\Pi_{\sigma}^{(\mu\nu)} \right]= d_{\sigma} m_{\sigma}^{(\mu\nu)}$, we
obtain the bound
\begin{eqnarray}\label{Bound}
\<c\>&\geq & a_{\sigma_0}+{\sum _{\mu,\nu,\sigma}}' ~a_{\sigma}~
m_{\sigma}^{(\mu\nu)}~|c_{\mu}c_{\nu}| \nonumber 
\\ &\equiv& \<c\>^{Opt}~.
\end{eqnarray}
It is straightforward to see that the choice of a covariant POVM with
$\Xi=|\eta\>\<\eta|$ with $|\eta\>$ given by (\ref{OptEta}) achieves
this lower bound.
\qed
\subsection{Remarks}\label{Remarks}
{\bf Remark I} Up to the constant term $a_{\sigma_0}$, the minimum
cost (\ref{OptCost}) is simply given by the expectation value of the \emph{cost
matrix}  (\ref{Matrix}) over the normalized vector
$\mathbf{v} \equiv (~|c_{\mu}|~)$. Therefore the optimal input state
is obtained just by finding the eigenvector corresponding to the
minimum eigenvalue of the cost matrix.  In other words, the
optimal state for the estimation of an unknown parameter is always a
superposition of maximally entangled states, with the coefficients in
the superposition modulated by the particular choice of the cost
function.  Notice the simplification of the optimization problem
provided by Theorem \ref{Theo2}: instead of optimizing a state in
the Hilbert space $\spc H= \bigoplus_{\mu \in
\set S}~ \spc H_{\mu}\otimes \Cmplx^{m_{\mu}}$ we need only to
optimize a vector in $\Reals^{|\set S |}$, where $|\set S|$ is the
number of irreducible representations contained in the action of the black box.

{\bf Remark II} The optimal POVM of Theorem \ref{Theo2} is the same optimal POVM
arising from the maximum 
likelihood criterion \cite{MLPovms,DeGiorgi}. In fact, this criterion
corresponds to the particular choice of the delta cost function
\begin{eqnarray*}
c({\hat g},g)&=&-\delta({\hat g},g)\\
&=&-\sum_{\sigma}~d_{\sigma}~\chi^{\sigma}({\hat g}g^{-1})~,
\end{eqnarray*} 
which is of the form (\ref{OurCBis}).  In other words, in the case of superpositions of maximally entangled states, the result of Theorem
\ref{Theo2} can be viewed as the extension of the maximum likelihood approach of Ref.  \cite{MLPovms}
to arbitrary cost functions. 

{\bf Remark III} In the optimization of covariant POVM's it is often
assumed that the operator $\Xi$ corresponding to an optimal estimation
can be taken with unit rank. However, for mixed states some
counterexamples are known \cite{ExtPovms,PhaseMixedStates}, and for pure states there is no general
proof that the POVM 
minimizing the average Bayes cost can be chosen with rank one. Therefore, it is important to emphasize  that here the rank-one
property of the optimal POVM of Theorem \ref{Theo2} is a result of the derivation, not an
assumption.\\
\section{Applications}\label{Applications}
\subsection{Optimal transmission of reference frames}\label{RefframeProof}
The result of Theorem \ref{Theo2} can be exploited to give the
definitive proof of optimality of the protocol for the absolute
transmission of a Cartesian reference frame of Ref.  \cite{refframe},
which concludes a long debate about the optimal way of communicating a
reference frame \cite{frames}. Such a protocol allows two distant
parties, Alice and Bob, to align their Cartesian axes in an absolute
way, i.e. without the need of any kind of prior information about
their relative orientations. To this purpose, Alice sends to Bob $N$
spin-$\frac{1}{2}$ particles, prepared in some fixed state. The
preparation procedure of the state is related to the directions of
Alice's Cartesian axes: for example Alice can align the angular
momenta of some particles with her $x-$axis, some with her $y-$axis,
and so on. When Bob receives the particles, since his axes are
mismatched with Alice's ones, each particle appears rotated by the
same unknown rotation. Then, instead of receiving the particles in the
same state prepared by Alice, Bob receives them in a rotated
state. Clearly, if he knows how the state should look in absence of
rotations, he can try to estimate the difference, i.e.  he can
estimate the unknown rotation, inferring in this way the directions of
Alice's axes.  The precision of this scheme is defined in a Bayesian
way, by taking as cost function the \emph{transmission error},
i.e. the distance between the directions of Alice's axes and Bob's
axes at the end of the protocol. In terms of the estimated rotation
${\hat g}$ and the true one $g$, the transmission error can be written
as \cite{refframe}
\begin{equation}\label{CostRefframe}
e(\hat{g},g)=6-2\chi^1(\hat g g^{-1})~,
\end{equation}
where $\chi^1(g)\equiv \Tr[U^1_g]$ is the character of the three
dimensional irreducible representation of the rotation group. It is
immediate to see that the transmission error is a cost function the
form (\ref{OurCBis}).

What is the best precision that can be achieved with the mentioned
protocol? To answer this question we need to solve two problems: the
first is to find what is the optimal state for encoding rotations, and
the second is to find the optimal estimation strategy.  It is
important to stress that, since we want to achieve an \emph{absolute}
transmission, we are not allowed to use an external reference system,
whose role would correspond to a partially shared reference frame
\cite{refframe}. For this reason we are allowed only to exploit the
entanglement coming from the multiple equivalent representations that
appear in the Clebsch-Gordan series of $U_g^{\otimes N}$, where $U_g$
is the $\SU 2$ matrix that represent the rotation $g$ in the
two-dimensional Hilbert space $\spc H$ of a single spin $\frac{1}{2}$
particle.

The tensor product Hilbert space $\spc H^{\otimes N}$ can be
decomposed as in (\ref{SpaceDecomp})
\begin{equation}\label{SpaceDecompSU2}
\spc H^{\otimes N}= \bigoplus_{j=0(\frac{1}{2})}^{\frac{N}{2}}~\spc H_j \otimes \Cmplx^{m_j}~.
\end{equation}
The irreducible representations are labeled by the quantum number $j$
of the total angular momentum, which ranges from $0(\frac{1}{2})$ to
$\frac{N}{2}$ for $N$ being even (odd), respectively. The dimension of
the representation space $\spc H_j$
is\begin{equation}\label{dimension} d_j=2j+1~,
\end{equation} 
while the multiplicities are given by \cite{CEM,refframe}:
\begin{equation}\label{multiplicity}
m_j=\frac{2j+1}{\frac{N}{2}+j+1}\binom{N}{\frac{N}{2}+j}~.
\end{equation}
Since $m_j \geq d_j$ for any $j<\frac{N}{2}$, it is possible to have
maximal entanglement between representation spaces and multiplicity
spaces for any $j$, with the only exception of $j=\frac{N}{2}$.

However, as shown in \cite{refframe}, the contribution of the subspace
with $j=\frac{N}{2}$ is negligible in the asymptotic limit of large
$N$.  Therefore we can restrict ourself to the subspace $\spc
H'=\bigoplus_{j=0(\frac{1}{2})}^{\frac{N}{2}-1}~\spc H_j \otimes \Cmplx^{m_j}$,
and consider the state
\begin{equation}\label{RefframeState}
|A\>\!\>= \bigoplus_{j=0(\frac{1}{2})}^{\frac{N}{2}-1}~
 \frac{c_j}{\sqrt{d_j}}|\openone_j\>\!\>~.
\end{equation}  
According to Theorem \ref{Theo1}, and to the result of
 \cite{AJV}, this is the optimal state in the subspace $\spc H'$ for
the estimation of an unknown $\SU 2$ rotation.  

Now we can use Theorem \ref{Theo2} to state that the optimal
estimation strategy is described by the covariant POVM given by
$\Xi=|\eta\>\<\eta|$ with
\begin{equation}\label{RefframePOVM}
|\eta\>=\bigoplus_{j=0 (\frac{1}{2})}^{\frac{N}{2}-1}~\sqrt{d_j}
|\openone_j\>\!\>~.
\end{equation} 
The optimization of the coefficients $c_j$ in the state
(\ref{RefframeState}) has been done in \cite{refframe}, where the POVM
(\ref{RefframePOVM}) was assumed by exploiting for simplicity the
maximum likelihood approach.  In this way, the results of Theorem
\ref{Theo1} and \ref{Theo2} provide the optimality proof for the
protocol proposed in Ref. \cite{refframe}. Therefore, we can
definitely state that the asymptotic precision
\begin{equation}
\<e\> = \frac{8\pi^2}{N^2}
\end{equation}
is the best that can be achieved for all input states and all POVM's,
namely it is the ultimate precision limit imposed by Quantum Mechanics
in the absolute alignment of two Cartesian reference frames.
\subsection{Optimal estimation of a completely unknown maximally
  entangled state}
\label{EntEstimation} Maximally entangled
states are a fundamental resource for quantum teleportation
\cite{teleport} and for quantum cryptography \cite{Ekert}. To achieve
ideal teleportation, Alice and Bob must know with precision which
maximally entangled state they are sharing, otherwise the fidelity of
the state received by Bob with the original state from Alice can be
lowered. Similar arguments apply to the cryptographic schemes where
the correlations arising from entanglement are exploited to generate a
secret key.

Now we will consider the problem of estimating in the best way a
completely unknown maximally entangled state, provided that $N$
identical copies are available.  This can be done as an application of
Theorem \ref{Theo2}.  Let's consider a state $|\psi\>\!\> \in \spc H
\otimes \spc H$, with $\dim (\spc H)=d$. In terms of the notation
(\ref{DoubleKet}), this state is maximally entangled if and only if
$\psi= \frac{1}{\sqrt{d}}~U$, where $U$ is some unitary
operator. Using property (\ref{DoubleKetProperty}), any maximally
entangled state can be written as
\begin{equation}\label{AnyEntState}
|\psi_g\>\!\>=\frac{1}{\sqrt{d}}~(U_g \otimes \openone) ~|\openone\>\!\>~,
\end{equation}
where $U_g$ is an element of the group $\SU d$.  

If $N$ identical copies of the unknown state $|\psi_g\>$ are given,
then the problem becomes to find the best estimate for parameter $g$
that labels the states of the form $|\Psi_g\>\!\>=|\psi_g\>\!\>^{\otimes
N}$. Optimality is defined here in terms of maximization of the
Uhlmann fidelity between the true state and the estimated one:
\begin{equation}
f({\hat g},g)= \left| \<\!\<\psi_{g}|\psi_{{\hat g}}\>\!\>\right|^2~.
\end{equation}
Using the definition (\ref{AnyEntState}) and the property (\ref{DoubleKetProduct}), we obtain 
\begin{equation}
f(\hat g,g)=\frac{1}{d^2}~|\chi({\hat g}g^{-1})|^2~.
\end{equation}where $\chi(g) =\Tr[U_g]$. The maximization of the fidelity corresponds to the minimization of the cost function
\begin{equation}\label{CostEntangled}
c({\hat g},g)=1-f({\hat g},g)~,
\end{equation}
which is of the form (\ref{OurCBis}).  In particular, for $d= 2$,
$|\chi (g)|^2=1 +\chi^1(g)$, where $\chi^1(g)=\Tr[U_g^1]$ is the
character of the irreducible representation of $\SU 2$ with angular
momentum $j=1$,whence we have
\begin{equation}
c({\hat g},g)= \frac{1}{4}~\left( 3 -\chi^1(g^{-1}{\hat g})\right)~.
\end{equation}

All the states of the form $|\Psi_g\>=|\psi_g\>\!\>^{\otimes N}$ are generated
from the input state
\begin{equation}\label{N-EntangledStates}
|\Psi\>=\frac{1}{\sqrt{d^N}}~|\openone\>\!\>^{\otimes N}
\end{equation} by the action of the
representation 
\begin{equation}\label{N-UTensorId}
\rep R(\grp G) =\{(U_g\otimes \openone)^{\otimes N}~|~ U_g \in
\SU d\}~.
\end{equation}  
Now we need to know how the input state is decomposed with respect to the invariant subspaces of this representation.
\begin{lemma}\label{DecompN-EntangledState}
Using suitable bases for the muliplicity spaces in decomposition (\ref{SpaceDecomp}), the input state (\ref{N-EntangledStates}) can be written  as
\begin{equation}
|\Psi\>=\bigoplus_{\mu \in \set S}~\frac{c_{\mu}}{\sqrt{d_{\mu}}}~|\openone_{\mu}\>\!\>~,
\end{equation}
where the sum runs over the irreducible representations of $\SU d$ occurring in the Clebsch-Gordan series of $\rep R (\grp G)$ (\ref{N-UTensorId}), and
\begin{equation}\label{CoeffN-EntangledState}
c_{\mu}=\sqrt{\frac{d_{\mu}m_{\mu}}{d^N}}~,
\end{equation}
$d_{\mu}$ and $m_{\mu}$ being respectively the dimension and the multiplicity of the representation
$\mu$ in the Clebsch-Gordan series of $\{U_g^{\otimes N}\}$. 
\end{lemma} \Proof See appendix.

Thank to this lemma we can exploit directly the result of Theorem
\ref{Theo2} to calculate the average fidelity.  Now we will carry on
the calculation of the optimal fidelity in the simplest case $d=2$.
As usual, the irreducible representations of $\SU 2$ are labeled by
the quantum number $j$, ranging from $0(\frac{1}{2})$ to $\frac{N}{2}$ for
$N$ being even (odd), respectively. The minimum cost can be evaluated
using Theorem \ref{Theo2} as
\begin{equation}
\<c\>^{Opt}= \frac{3}{4}+ \sum_{i,j=0(\frac{1}{2})}^{\frac{N}{2}} ~ |c_i| ~C_{ij}~ |c_j|
\end{equation}
Using Eq. (\ref{CoeffN-EntangledState}) with the values of dimensions and
multiplicities given by (\ref{dimension}) and (\ref{multiplicity}), the coefficients of the state become        
\begin{equation}\label{StateCoeff}
c_i= g(i)~\sqrt{\frac{1}{2^N}\binom{N}{\frac{N}{2}+i}}~,
\end{equation}
where
\begin{equation}
g(i)= \frac{2i+1}{\sqrt{\frac{N}{2} +i+1}}~.
\end{equation}
On the other hand,  the matrix $C_{ij}$ is
calculated  according to the definition (\ref{Matrix}), namely by evaluating the multiplicity of the
representation with angular momentum $k=1$ in the Clebsch-Gordan
series of the tensor product $U^i_g\otimes U^{j*}_g$~. In this way we get
\begin{eqnarray}
C_{ij}&=& -\frac{1}{4} (
\delta_{i,j} + \delta_{i,j+1} + \delta_{i,j-1})~.
\end{eqnarray}
Since $\sum_i |c_i|^2=1$, we have
\begin{equation}\label{OptCostSU2}
\<c\>=\frac{1}{2}\left(1 - \sum_{j=0(\frac{1}{2})}^{\frac{N}{2}-1}~ c_j c_{j+1}\right)~.
\end{equation}
To obtain the asymptotic behavior of the optimal fidelity, we can
approximate the binomial distribution in (\ref{StateCoeff}) with a
Gaussian $G_{\sigma}(x)$ with mean $\bar x=0$ and variance
$\sigma^2=\frac{N}{4}$.  Since the sum in (\ref{OptCostSU2}) runs over
a large interval with respect to $\sigma$, we can also approximate it
with an integral over $[0,+\infty]$~. All these approximations hold up to
order higher than $\frac{1}{N}$. Thus the evaluation of the optimal fidelity is reduced to the evaluation of the integral
\begin{equation}
I=\int_0^{\infty} \d x ~ g(x)g(x+1)~ G_{\sigma}(x)~,
\end{equation}
whose leading order can be obtained from Taylor expansion.
In this way, we derive the asymptotic cost
\begin{equation}
\<c\>^{Opt}=\frac{3}{4N}~,
\end{equation}
corresponding to the optimal fidelity
\begin{equation}\label{OptFidEnt}
\<f\>^{Opt}=1-\frac{3}{4N}~.
\end{equation}
Remarkably, the Bayes cost with uniform a priori distribution has the
same asymptotic behavior of the cost of the optimal locally unbiased
estimator obtained in \cite{Ballester}, for any possible value $g$ of
the true parameter. This means that in the present unbiased case the covariant measurement of
Theorem \ref{Theo2} is optimal not only on average but also pointwise.

\section{Conclusions}

In this paper we solved the general problem of optimal estimation of
group transformations in the Bayesian framework with uniform prior.
For this purpose, we introduced a class of cost functions generalizing
the Holevo class for phase estimation, containing the maximum
likelihood strategy as a special case.  For this family of cost
functions, we derived the general form of the optimal input states,
which involves maximal entanglement between representation and
multiplicity spaces of the group action. More precisely, the form of
an optimal input state is a direct sum of maximally entangled states,
and for a given cost function one only needs to optimize the
coefficients in the sum. Moreover, \emph{for any state of the optimal
form all invariant cost functions lead to the same optimal POVM}. In
this way, it is possible to derive an explicit expression for the
average cost and to reduce the optimization of the state to a simple
eigenvalue problem.  As applications of the general result we have
given the first complete derivation of the ultimate precision limit
imposed by Quantum Mechanics in the absolute alignment of two
Cartesian reference frames, and we have derived the optimal estimation
of a completely unknown two-qubit maximally entangled state with $N$
copies of the state. In the present paper we focused attention to
compact groups and finite dimensional Hilbert spaces, nevertheless an
extension of our results to infinite dimensional Hilbert spaces and
non-compact groups is possible, in the same way as in
\cite{DeGiorgi}. However, since in infinite dimension the optimal
states may be non-normalizable, one has to approximate them with
physical states by fixing additional constraints as, typically, the
energy constraint.

\acknowledgments This work has been supported by the FET European
Networks on Quantum Information and Communication Contract
IST-2000-29681:ATESIT, by MIUR 2003-Cofinanziamento, and by INFM
PRA-2002-CLON.
\section{Appendix}
\subsection{Invariant cost functions}
In this section we prove the form (\ref{OurC}) of any invariant cost function.
\begin{prop}\label{Ergodic}
The following integral formula holds:
\begin{equation}
\int \d g~ U_g^{\mu} \otimes U_g^{\nu~*} =\delta_{\mu\nu}~ \frac{|\openone_{\mu}\>\!\>\<\!\<\openone_{\mu}|}{d_{\mu}}~.
\end{equation}
\end{prop}
\Proof Using Eq. (\ref{IntegralProject}), 
we recognize in the l.h.s. the projection onto the subspace of $\spc
H_{\mu} \otimes \spc H_{\nu}$ that carries the trivial representation
in the Clebsch-Gordan decomposition of $U_g^{\mu} \otimes
U_g^{\nu~*}$. Using the orthogonality of characters, one can prove
that such tensor product contains the trivial representation if and
only if $\mu=\nu$. Moreover, if $\mu=\nu$, then the multiplicity of the
trivial representation is one. Using the property
(\ref{DoubleKetProperty}), we see that the vector
$|\openone_{\mu}\>\!\>$ is invariant under $U_g^{\mu} \otimes
U_g^{\mu~*}$. Therefore the r.h.s. is the projection onto the
one-dimensional invariant subspace that carries the trivial
representation, whence it coincides with the l.h.s. \qed
\begin{prop}
Any invariant function $c(\hat g, g)$ has the form
\begin{equation}
c(\hat g, g)=\sum_{\mu}~a_{\mu}~ \chi^{\mu}(g^{-1}\hat g)~,
\end{equation}
where $\chi^{\mu}(g)\equiv \Tr[U_g^{\mu}]$.  
\end{prop}
\Proof   For each irreducible representation $\mu$,
 consider the matrix elements $u_{ij}^{\mu}(g) \equiv \<\psi^{\mu}_i|~U_g^{\mu}~|\psi^{\mu}_j\>$ with respect to a fixed basis
 $\Base^{\mu}=\{|\psi^{\mu}_i\>~|~i=1, \dots, d_{\mu}\}$ for the
 representation space $\spc H_{\mu}$.  Since the collection of all
 these matrix elements is an orthogonal basis for $L^2(\grp G)$ \cite{groups}, we
 can expand the function $c(\hat g,g)$ as
\begin{equation}
c(\hat g,g)=\sum_{\mu,\nu}\sum_{i,j=1}^{d_{\mu}}\sum_{k,l=1}^{d_{\nu}}~a_{ijkl}^{\mu\nu}~ u_{ij}^{\mu}(\hat g)~u_{kl}^{\nu*}(g)~,
\end{equation}
where the complex conjugate in $u_{kl}^{\nu *}(g)$ is for later
convenience.  Now, the function $c$ is both left- and right-invariant,
whence it coincides with its average $\bar c(\hat g, g)\equiv \int \d
k \int \d h ~ c(k\hat gh,kgh)$. Using Proposition \ref{Ergodic} and
Eqs. (\ref{DoubleKetProduct}) and (\ref{DoubleKetProperty}), we obtain
\begin{equation*}
\begin{split}
c(\hat g,g)&=\int \d k \int \d h~ c(k\hat g h,kg h)\\
&=\sum_{\mu,\nu}\sum_{i,j,k,l}~a^{\mu\nu}_{ijkl}~\delta_{\mu\nu}
~\times\\ &\times 
\<\psi^{\mu}_ i|\<\psi^{\mu}_ k|
\frac{|\openone_{\mu}\>\!\>\<\!\<\openone_{\mu}|}{d_{\mu}}(U^{\mu}_{\hat
g}\otimes U^{\mu
*}_{g})\frac{|\openone_{\mu}\>\!\>\<\!\<\openone_{\mu}|}{d_{\mu}}|\psi^{\mu}_
j\>|\psi^{\mu}_ l\>~~~~~\\
&=\frac{1}{d_{\mu}^2}~\sum_{\mu}\sum_{i,j,l,k} a^{\mu\mu}_{ijkl}~
\delta_{ik}~\delta_{jl}~\Tr[~U_{\hat g g^{-1}}^{\mu}~]\\ &=\sum_{\mu}
~a_{\mu}~ \chi^{\mu}(g^{-1} \hat g)~,
\end{split}
\end{equation*}
where $a_{\mu} \equiv
\frac{1}{d_{\mu}^2}\sum_{i,j}~a_{ijij}^{\mu\mu}$.\qed

\subsection{Decomposition of a product of maximally entangled states}

Here we give the proof for Lemma \ref{DecompN-EntangledState}.  \Proof
Consider the representation $\rep R (\grp G)=\{(U_g \otimes
\openone)^{\otimes N}~|~ U_g \in \SU d\}$.  It is convenient to order
the $2N$ Hilbert spaces in the tensor product $\spc H^{\otimes 2N}$ in
such a way that the unitary operators act on the first $N$ spaces,
while the identity operators acts on the second $N$ spaces. With this
ordering, by defining $\spc H_A$ ($\spc H_B$) the tensor product of
the first (second) $N$ spaces, we have $\rep R (\grp G)=\rep R_A (\grp
G) \otimes \openone_B$, where $\rep R_A (\grp G)\equiv\{U_g^{\otimes
N}~|~U_g \in \SU d\}$ is the $N$-fold tensor representation of $\SU
d$.

Let's decompose now the Hilbert space $\spc H_A$ with respect to the
action of the representation $\rep R_A (\grp G)$:
\begin{equation}\label{R-ASpaceDecomp}
\spc H_A =\bigoplus_{\mu}~ \spc H_\mu \otimes \Cmplx^{m_{\mu}}~.
\end{equation}
The tensor product $\spc H_A \otimes \spc H_B$ can be decomposed with
respect to $\rep R (\grp G)=\rep R_A (\grp G) \otimes \openone_B$ as
\begin{equation}\label{SpaceDecompForMaxEnt}
\spc H_A \otimes \spc H_B = \bigoplus_{\mu}~ \spc H_{\mu} \otimes \Cmplx^{M_{\mu}}~,
\end{equation}
where the multiplicity has been increased to $M_{\mu}=m_{\mu} \times
d^N$, since $\spc H_B$ has been absorbed in the multiplicity spaces.

With respect to the factorization $\spc H^{\otimes 2N}=\spc H_A
\otimes \spc H_B$, the input state $|\Psi\>=|\openone\>\!\>^{\otimes
N}$ can be written as
\begin{equation}
|\Psi\>=\frac{1}{\sqrt{d^N}}~|\openone^{\otimes N}\>\!\>~,
\end{equation}
where $\openone^{\otimes N}$ is the identity in $\spc H_A\equiv \spc
H^{\otimes N}\equiv \spc H_B$. Here we are using notation
(\ref{DoubleKet}), with respect to the product basis $\Base^{\otimes
N}$ for $\spc H_A$ and $\spc H_B$, $\Base$ being a fixed basis for
$\spc H$.  Now we want to change the basis in $\spc H_A$, by switching
from $\Base^{\otimes N}$ to $\Base'\equiv
\bigcup_{\mu}~\Base^{\mu}_{R}\otimes \Base_{M}^{\mu}$, where
$\Base^{\mu}_{R}\equiv \{|\psi^{\mu}_n\>~|~n=1, \dots ,d_{\mu}\}$
($\Base^{\mu}_{M}\equiv \{|\phi^{\mu}_n\>~|~n=1,\dots ,m_{\mu}\}$) is
a basis for the representation (multiplicity) space in
Eq. (\ref{R-ASpaceDecomp}). One has
\begin{equation*}
\begin{split}
|\Psi\>&=\frac{1}{\sqrt{d^N}}~\bigoplus_{\mu}~\sum_{m=1}^{d_{\mu}}~\sum_{n=1}^{m_{\mu}}~|\psi^{\mu}_m\>_A|\phi^{\mu}_n\>_A~|\psi^{\mu
  *}_m\>_B|\phi^{\mu *}_n\>_B\\
&=\frac{1}{\sqrt{d^N}}~\bigoplus_{\mu}~\sum_{m=1}^{d_{\mu}}~\sqrt{m_{\mu}}~
|\psi^{\mu}_m\>~|\tau^{\mu}_m\>~,
\end{split}
\end{equation*} 
where we defined the normalized vector
\begin{equation}
|\tau^{\mu}_m\>\equiv \frac{1}{\sqrt{m_{\mu}}}~\sum_{n=1}^{m_{\mu}}~ |\phi^{\mu}_n\>|\psi^{\mu *}_n\>|\phi^{\mu *}_n\>~.
\end{equation}
Therefore, exploiting notation (\ref{DoubleKet}) with respect to the
bases $\{|\psi_m^{\mu}\>\}$ and $\{|\tau^{\mu}_m\>\}$, we can write
\begin{equation}
|\Psi\>=\bigoplus_{\mu}~\sqrt{\frac{m_{\mu}}{d^N}}~|\openone_{\mu}\>\!\>~.
\end{equation}
\qed

\end{document}